\documentclass{article}

\usepackage[dvips]{graphicx}

\title{Unusual scenario of the temperature evolution of magnetic state in novel carbon-based nanomaterials}
\author{V. A. Ryzhov$^{1,2}$, A. V. Lashkul$^{2}$, V. V. Matveev$^{3}$,  M. V. Mokeev$^{4}$,\\ P. L. Molkanov$^{1}$,  A. I. Kurbakov$^{1,3}$, K. G. Lisunov$^{2,6}$, I. A. Kiselev$^{1}$,\\ D. Galimov$^{2,5}$,  E. L\"ahderanta$^{2,7}$}

\begin{document} 
\maketitle
\noindent\thanks{${}^{1}$Petersburg Nuclear Physics Institute, NRC ``Kurchatov Institute'', Orlova Coppice, Gatchina, Leningrad province, 188300, Russia \\
${}^{2}$Department of Mathematics and Physics, Lappeenranta University of Technology, PO Box 20, FIN-53851 Lappeenranta, Finland \\
${}^{3}$Saint-Petersburg State University, Petergof, Ulyanovskaya str., 1, St. Petersburg, 198504, Russia \\
${}^{4}$Institute of Macromolecular Compounds, Russian Academy of Sciences, St. Petersburg, Russia \\
${}^{5}$South Ural State University, 454080 Chelyabinsk, Russia}\\
${}^{6}$Institute of Applied Physics ASM, Academiei Str., 5, MD 2028 Kishinev, Moldova

\footnotetext [7] {Corresponding author; e-mail:Erkki.lahderanta@lut.fi}
\begin{abstract}
Two porous carbon-based samples doped with Au and Co, respectively, are investigated. The neutron diffraction study reveals an amorphous structure of both samples. The Co-doped sample contains structural clusters with larger size and exhibits a long-range ferromagnetic (FM) ordering at 2.6 K. The NMR investigations demonstrate, that the samples are obtained with a partial carbonization of initial aromatic compounds and do not reach a state of glassy carbon. The magnetization study, as well as investigations of a longitudinal nonlinear response to a weak \textit{ac} field and electron magnetic resonance, gives evidences for presence of FM clusters in the samples already at temperatures well above 300 K. A short-range character of the FM ordering in the Au-doped sample transforms below \textit{T}${}_{C }$$\approx$ 210 K into another inhomogeneous FM state. Besides the FM clusters, this state contains a subsystem with a long-range FM ordering (matrix) formed by paramagnetic centers, existing outside the clusters. The nonlinear response data suggest a percolative character of the long-range FM matrix, which is connected probably with a porous sample structure. The magnetization data give evidence for the formation of an inhomogeneous state in the Co-doped sample, similar to that in the Au-doped one. However, this state is formed at higher temperatures, lying well above 350 K, and exhibits a more homogeneous arrangement of the FM nanoparticles and the FM matrix. Temperature dependence of the magnetization in the Au-doped sample, measured in low field after cooling in zero magnetic field, exhibits a transition from a slow increase to a fast decrease at \textit{T} $\sim$ 140 K, lying below \textit{T}${}_{C}$. This is attributable to changes of the domain formation regime in the FM matrix on cooling, connected with the inhomogeneous character of the FM state in the Au-doped sample. Such peculiarity of the magnetization behavior is absent in the Co-doped sample below 350 K, which is in agreement with formation of the FM state in this sample at much higher temperatures. Further cooling below \textit{T} $\sim${} 3(10) K leads to a steep increase of the magnetization in the sample doped with Au (Co). This is attributable to the domain rearrangement in the inhomogeneous FM state at low temperatures. The distinctions in the nonlinear response to a weak \textit{ac} field, observed in different parts of the Co-doped sample, indicate existence of a large-scale spatial magnetic inhomogeneity in the initial sample. 
\end{abstract}

\section{Introduction}
 Nowadays, it is evident that carbon-based nanomaterials represent a novel class of ferromagnetic (FM) matter, which does not contain basically any FM metal components [1]. Such materials attract considerable attention due to a high-temperature FM behavior observed in various carbon structures, accompanied with magnetic hysteresis and the remanent magnetisation. These features make the materials above quite attractive for applications not only in technique, but in biology and medicine as well, which is connected with low toxicity due to vanishing concentration of metallic elements [1, 2]. 

Experimental investigations establishing intrinsic magnetism of defect-rich carbon structures [1, 2] have been supported by extensive theoretical work. Namely, the FM behavior has been predicted in such structures as (i) graphite surface with negative Gaussian curvature [3]; (ii) a mixture of carbon atoms with alternation of \textit{sp}${}^{2}$ -- \textit{sp}${}^{3}$ bonds [4]; (iii) those containing the graphene zigzag edges [5, 6]; (iv) disordered graphite with random single-atom defects [7]. In turn, theoretical values of a local magnetic moments $\mu $ $\sim$ 1 {--} 2 $\mu $${}_{B}$, connected with intrinsic defects or disorder [8, 9], have been supported by experimental investigations of highly-oriented pyrolitic graphite yielding $\mu $ $\sim$ 0.2 {--} 1.5 $\mu $${}_{B}$ per defect at a distance between defects $\sim$ 0.5 {--} 4 nm [10].

Magnetic properties of powder and glassy carbon-based nanomaterials, including those doped with metals, have been investigated recently [11{--}13], remaining, however, many important details unclear. These include a structure and a local structure of the compounds, a character of the FM ordering, a possible similarity of the local structure to that of fullerenes or other carbon-based materials and, eventually, distribution of the magnetization over the sample volume. 

In this work, we have studied two carbon-based compounds doped by Au and Co (note that Au is nonmagnetic, whereas Co ions may possess a local magnetic moment), which are similar to the Au-doped and Co-doped samples, respectively, investigated in [11{--}13]. We have obtained information on their structural and magnetic properties, using several independent methods to clarify some of the issues mentioned above. Magnetic behavior of the Au-doped sample gives evidence for presence of a short-range FM order near the room temperature. Taking into account the observed amorphous structure of the Au-doped sample, this implies formation of FM nanoparticles at higher temperatures in agreement with the results of [11{--}13]. However, a subsystem of paramagnetic (PM) centers (referred below as "matrix"), not involved in formation of the FM clusters, have been found in the Au-doped sample on cooling. This matrix exhibits a magnetic ordering, leading eventually to an inhomogeneous FM state in the Au-doped sample. The obtained results suggest formation of such a state in the Co-doped sample at much higher temperatures. At low temperatures, a complex temperature evolution of the FM spin arrangement of this state, accompanied with appearance of a long-range FM order, is observed in both samples. The distinctions in the nonlinear response to a weak \textit{ac} field, measured in different large-scale parts of the initial Co-doped sample, suggest corresponding differences of their magnetic state, as well as of the properties and density of the FM nanoparticles entering them.

\section{Experimental details}

The porous glassy carbon samples doped with 0.004 mass \% of Au (S--Au) and with 0.117 mass \% of Co (S--Co), which have been prepared and studied earlier in [11--13], were investigated.  The preparation details have been described in Refs. 11 -- 13. The atomic force microscopy (AFM) investigations of the samples doped with Ag, Au and Co gave evidence for the presence of carbon particles with a broad size distribution given by the average, \textit{R}${}_{av}$, and the maximum, \textit{R}${}_{max}$, particle radii [11--13]. The values of \textit{R}${}_{max}$ $\sim${} 60 nm and \textit{R}${}_{av}$ $\sim${} 110 nm were found close in all the samples above.

In this work the structure and the magnetic state of the samples were studied with neutron diffraction, whereas the local structure was obtained with solid-state NMR investigations. Magnetic properties were investigated by measurements of the \textit{dc} magnetization with a SQUID magnetometer, by registration of the second harmonic of a longitudinal nonlinear response to a weak \textit{ac} magnetic field and by recording the electron magnetic resonance spectra.

The neutron powder diffraction study was carried out using the PNPI superpositional diffractometer equipped with 48 counters in four sections at WWR-M reactor, beam line 9. Neutron diffraction patterns were measured at 2.6 and 300 K in the superposition mode, using monochromatic neutrons with a wave length $\lambda $ = 1.7526 \AA {} in the angular range of 4${}^\circ$ $\leq$ 2$\theta$ $\leq$ 145${}^\circ$ with a step of 0.1${}^\circ$. 

Solid-state NMR spectra were recorded under magic angle spinning (MAS) conditions at ambient temperature, using the spectrometer AVANCE II-500WB (Bruker) operated at 125.8 MHz for ${}^{13}$C nucleus. Samples were packed in a 4 mm zirconium rotor and spun at a 7 kHz frequency. Single-contact ${}^{1}$H$\rightarrow^{13}$C cross-polarization (CP) technique with 3 ms contact time was applied for ${}^{13}$C CP-spectra recording with high power proton decoupling at frequency of 100 kHz. To increase the CP-efficiency, a sample was blended with proton containing chemical inert material (Al(OH)${}_{3}$) in a 1:1 weight composition. For ${}^{13}$C direct polarization spectrum, 2.3 $\mu $s pulse ($\pi $/4) was used with repetition time of 6 s and proton decoupling at frequency of 100 kHz. The number of scans was 4k or 8k to obtain a reasonable signal-to-noise ratio. All chemical shifts are given in ppm from tetramethylsilane. Deconvolution of the obtained spectra was performed using the DMFIT software [14].

Magnetization, \textit{M} (\textit{B}), was measured with a SQUID magnetometer in the magnetic field \textit{B }up to 5 T by increasing and decreasing the field. The dependence of \textit{M} (\textit{T}) was measured in a constant magnetic field \textit{B} between 1 mT -- 5 T, after cooling the sample from 300 K down to 3 K in a zero field (zero-field cooled magnetization, \textit{M}${}_{ZFC}$) or in the applied field (field-cooled magnetization, \textit{M}${}_{FC}$). Thermoremanent magnetization (TRM) was investigated after cooling the sample from 300 K to 3 K in a non-zero magnetic field and reducing the field to zero. The magnetization data are presented below after subtraction of the diamagnetic contribution.

The measurements of the second harmonic of the magnetization, \textit{M}${}_{2}$, of the longitudinal nonlinear response were performed in the parallel steady and alternating magnetic fields, \textit{H }(\textit{t}) = \textit{H} + \textit{h }sin{ }$\omega$\textit{t} (where \textit{h} $\approx$ 14.3 Oe and \textit{f} = $\omega$\textit{/}2$\pi$ $\approx$ 15.7 MHz) under the condition of \textit{M}${}_{2}$ $\propto$ \textit{h}${}^{2}$. The latter permits us to analyze the results in frameworks of the perturbation theory. The real and imaginary phase components of \textit{M}${}_{2}$, Re \textit{M}${}_{2}$ and Im \textit{M}${}_{2}$ respectively, were recorded simultaneously as functions of \textit{H} at different sample temperatures between 100 and 350 K. The field \textit{H} was scanned symmetrically with respect to the origin to control the magnetic field hysteresis of the signal. According to symmetrical properties of \textit{M}${}_{2}$, its presence in the response is connected with existence of a spontaneous FM moment in a sample. The amplitude of \textit{H} was 300 Oe. The installation and the method of separation of the \textit{M}${}_{2}$-phase components are described in [15]. The sensitivity of the measurements above was $\sim${} 10${}^{-9}$ emu. The applied method permits us to detect formation of the FM clusters in a PM media and to trace temperature evolution of the cluster subsystem, due to its extreme sensitivity to the FM fraction of a sample. This has been demonstrated by our investigations of the Nd${}_{1-x}$Ba${}_{x}$MnO${}_{3}$ and La${}_{0.88}$MnO${}_{x}$ manganite perovskites [16, 17]. 

 Measurements of the electron magnetic resonance (EMR) spectra were performed with a special home-made X-range (\textit{f} = 8.37 GHz) ESR spectrometer, which provided a high-sensitive registration of wide lines [18]. It is supplied with a cylindrical two-mode balanced cavity with H${}_{111}$ type of the electromagnetic oscillations. The steady magnetic field \textbf{H} was directed along the cylinder axis (\textit{z} axis). A sample was placed at the bottom of the cavity, where it was acted by linearly polarized\textit{ ac}-field \textbf{h} directed along \textit{x} axis produced by excitation mode. The recorded signal in receiving mode was proportional to the off-diagonal component of the magnetic susceptibility tensor, \textit{M${}_{y }$}($\omega$) = $\chi$\textit{${}_{yx}$}($\omega$)\textit{h${}_{x}$}($\omega$).

\section{Results and discussion}

\subsection{ Neutron powder diffraction}

\noindent The diffraction patterns for both investigated samples at room temperature and for the Co-doped sample at \textit{T} = 2.6 K are displayed in Fig. 1. It can be seen that the lines for the Co-doped sample are appreciably narrowed. This suggests a larger size of structural clusters formed in the Co-doped sample with respect to those of the Au-doped sample. At the same time, positions of the main maxima are almost identical for both samples, implying a close internal cluster structure. For the Co-doped sample, the peak amplitudes increase with cooling from 300 K down to \textit{T} = 2.6 K, as follows from the differential signal (curve 4). This testifies to presence of the magnetic scattering, connected with a long-range FM order in the Co-doped sample. The effect of the temperature factors on the neutron diffraction patterns is much smaller than the observed difference.
 
\subsection{NMR with magic-angle spinning.}

 For better understanding of the composition and molecular structure of the samples, we obtained the NMR spectra of hydrogen (${}^{1}$H) and carbon (${}^{13}$C) nuclei under the MAS conditions. The main peaks in ${}^{1}$H spectrum (see Fig. 2) correspond to the aliphatic (2.3 ppm) and the aromatic (6.8 ppm) groups. The signal arising from the aromatic groups exhibits considerable anisotropy of the chemical shift (CSA), expectable for the aromatic protons. The aromatic to aliphatic signal ratio  $\sim$ 4 was found.

 The ${}^{13}$C{ }CPMAS spectrum displayed in Fig. 3 reveals four main lines, which can be interpreted as follows: (i) the 126 ppm line is connected with polynuclear aromatics, and a half-width of the peak, constituting only 15 ppm, looks quite narrow for such systems, which points out to a high isotropy of an environment; (ii) the 137 ppm line is attributed to substituted aromatics with a bent structure; (iii) the 153 ppm lineis connected with an oxygen-substituted aromatic, belonging probably to carbonyls; (iv) the 36 ppm line is attributed to aliphatic chains (corresponding to the peak at 2 ppm of the ${}^{1}$H spectrum in Fig. 2).

 Using CPMAS spectrum at lower spin frequency (4.5 kHz) [19], we obtained main components of the CSA tensor of the 126 ppm line: $\delta{}_{11}$ = 219.17, $\delta{}_{22}$ = 145.18 and $\delta{}_{33}$ = 11.25. Such magnitudes are typical of the carbon atoms with the \textit{sp}${}^{2}$-type of a chemical bond in the aromatic compounds (such as benzene, graphite etc.), which confirms the aromatic origin of the investigated material. In addition, the obtained spectra permit us to conclude, that the material has no fullerene fragments, because no lines have been detected either at 143 ppm (corresponding to C$_{60}$ with narrow characteristic line) or within the range of 130 -- 150 ppm (corresponding to C${}_{70}$ and higher fullerenes). 

To estimate the composition of the material, a direct-acquisition spectrum with proton decoupling (hpdec) has been obtained. The separation of the obtained signal into the spectral components permits us to obtain their integral intensities and thus to estimate contributions of different structural fragments to the composition of the investigated material (see Table 1).

To summarize, the NMR data demonstrate that the investigated material consists of the partially polymerized aromatic compounds (or polynuclear aromatics). Because the initial materials for synthesis of the glassy carbon samples were just aromatic compounds of various types (the furfuryl alcohol, ether iso-octylphenol, and dibutyl phthalate) [11--13], it is natural to conclude that the basis of the final material is derived of the partially transformed initial components. It is worth mentioning, that the spectra are very similar to those of the products of carbonization of the polyfurfuril alcohol at an incomplete heat treatment [20], but do not correspond completely to the spectrum of the commercial glassy carbon [21].

According to the NMR data, our samples are composed of the aromatic and aliphatic organic fragments. Their random distribution, typical of the amorphous structure, is found from the neutron diffraction investigations. This implies presence of multiple intrinsic defects acting as PM centers, which are well stabilized in aromatics and govern their magnetic properties.

\subsection{ Second harmonic of magnetization of the longitudinal nonlinear response}

The field and temperature dependences of the phase components, Re  \textit{M}${}_{2}$(\textit{H, T}) and Im   \textit{M}${}_{2}$(\textit{H, T}), of the second harmonic of magnetization, \textit{M}${}_{2}$, of the longitudinal nonlinear response were obtained in the  regimes of slow cooling and slow heating of a sample. Stabilization time of the sample temperature before the signal recording was not less than 300 s at any \textit{T}. Because the neutron diffraction data discussed in \textit{Subsection 1} indicate presence of the long-range FM order at least in S-Co, we use samples in a form of a plate, orienting the magnetic field \textit{H }(\textit{t}) along its plane, to decrease a possible demagnetization effect. A typical in-plane size exceeds the thickness of the plate by more than five times for both samples. The dependences of Re  \textit{M}${}_{2 }$(\textit{H}) and Im  \textit{M}${}_{2 }$(\textit{H}) for the Au-doped sample at several characteristic temperatures are displayed in Fig. 5. At \textit{T} = 293.3 K, both phase components of \textit{M}${}_{2}$ reveal a typical signal from FM clusters with an extremum in a weak steady field \textit{H}${}_{ext }$$\approx$ 20 Oe, as well as the field hysteresis. The latter resembles the hysteresis in doped manganites, where the FM clusters are formed already above \textit{T}${}_{C}$ due to the magneto-electronic phase separation [16, 17]. With lowering the temperature down to 235.5 K, another signal is added to the FM cluster signal of the Re \textit{M}${}_{2}$ component, exhibiting the dependence on \textit{H} close to linear one (see Fig. 5 (a), panel 235.5 K). Such signal is typical of a PM matrix in the critical regime.This suggests that a part of the PM centers is not involved into the FM clusters during their formation, but exhibits a short-range exchange interaction. We attribute such centers to the matrix. Therefore, the sample can be characterized by a state of the magnetic phase separation. 

When the second-order susceptibility can be introduced under the condition of \textit{M}${}_{2}$ $\propto$ \textit{h}${}^{2}$, the response of exchange magnets in the PM region is given by the expression,

\begin{equation} \label{GrindEQ__1_} 
\chi _{2} (\omega )=\frac{\Gamma }{-2i\omega +\Gamma } \chi _{2} -i\omega \frac{(\partial /\partial \omega _{0} )\Gamma }{(-2i\omega +\Gamma )(-i\omega +\Gamma )} \chi . 
\end{equation} 
Here,

\begin{equation} \label{GrindEQ__2_}
\chi _{1} =\chi _{1} (\tau ,H)=\partial M(\tau ,H)/\partial H \textnormal{ and }\chi _{2} =\chi _{2} (\tau ,H)=\frac{1}{2} \partial ^{2} M(\tau ,H)/\partial H^{2},       
\end{equation}

\noindent ${\omega}_{0}$ = \textit{g}$\mu$${}_{B}$\textit{H/$\hbar$} is the Larmor frequency and $\tau$\textit{= }(\textit{T -- T}${}_{C}$)/\textit{T}${}_{C}$ [22]. The first term in the right-hand side of Eq. (\ref{GrindEQ__1_}) is connected with a nonlinearity of the magnetization curve, \textit{M }(\textit{H}), and the second term to influence of the external magnetic field on relaxation processes. When the latter is absent, one has $\partial\Gamma/\partial\omega_{0} = 0$ and the second term vanishes. Eq. (\ref{GrindEQ__1_}) can be applied also to the analysis of the \textit{M}${}_{2}$ data (obtained under the condition of \textit{M}${}_{2}$${}_{ }$$\propto$ \textit{h}${}^{2}$) for an ensemble of single domain (SD) magnetic particles in a superparamagnetic (SP) regime [23]. A hysteretic response of\textit{M}${}_{2 }$(\textit{H}) indicates presence of a strong magnetic anisotropy, characterized by the inequality \textit{KV $>$ k}${}_{B}$\textit{T} at \textit{T $<$ T}${}_{b}$, where \textit{K} is the anisotropy constant, \textit{V} is the particle volume and \textit{T}${}_{b}$ is the blocking temperature. However, in such a case Eq. (\ref{GrindEQ__1_}) can be used only for a qualitative analysis of the \textit{M}${}_{2}$ data. Under a weak-field limit of \textit{g}$\mu$\textit{H $<$$<$} $\Omega$($\tau$) = \textit{kT}${}_{C}$$\tau^{5/3}$, where W is the energy of critical fluctuations, one obtains the expressions 

\begin{equation} \label{GrindEQ__3_}
 \chi _{1} (\tau )\propto \tau ^{-\gamma } \textnormal{ and Re } M_{2} (\tau )\propto H\tau ^{-\gamma _{2} } . \end{equation}                                         

\noindent Here, $\gamma$ and $\gamma$${}_{2}$ are critical exponents of the linear susceptibility and of \textit{M}${}_{2}$, respectively, with the values of $\gamma$ { = }4/3 and ${\gamma}_{2}$ { = }14/3, predicted for a cubic ferromagnet. Therefore, Re  \textit{M}${}_{2 }$(\textit{H},{ }$\tau$) is characterized by a linear dependence on \textit{H} in the PM region with Re  \textit{M}${}_{2 }$(\textit{H}=0) = 0. The appearance of the hysteretic signal with Re  \textit{M}${}_{2 }$(\textit{H}=0) $\neq$ 0 indicates a presence of the remanent magnetization, which is related to formation of a spontaneous FM moment in a sample. Note, that the \textit{M}${}_{2 }$ response of the cubic ferromagnet CdCr${}_{2}$Se${}_{4}$ in the critical PM region, 2\textit{T}${}_{C}$${}_{ }$\textit{$>$ T $>$ T}${}_{C}$, is well described by Eq. (\ref{GrindEQ__1_}) [22]. Crossover from the weak-field limit above to the strong-field limit, \textit{g}$\mu$\textit{H }$>$$>$ $\Omega$($\tau$), is accompanied with appearance of an extremum in the dependence of Re  \textit{M}${}_{2 }$(\textit{H}). The position of this extremum, \textit{H}${}_{ext}$, is shifted towards low fields with decreasing \textit{T}, because the energy of critical fluctuations, $\Omega$, depends on $\tau$, reaching a minimum at ${T}_{C}$. At the same time, the extremum amplitude is increased with lowering the temperature, exhibiting a maximum near \textit{T}${}_{C}$. Below \textit{T}${}_{C}$ but before the onset of the domain formation, \textit{H}${}_{ext}$${}_{ }$is shifted towards strong fields. Hence, the dependences of \textit{H}${}_{ext }$ and Re   \textit{M}${}_{2ext }$ on \textit{T} exhibit extremum near \textit{T}${}_{C}$, indicating a qualitative similarity to a second-order magnetic transition. To control the magnetic state of the samples, below we use the temperature dependences of signal parameters, which are presented in Fig. 7 (b). 

As follows from the plots at 235.5 K and 214.7 K in Fig. 5 (a), the second extremum (minimum at \textit{H} $>$ 0) appears in the Re \textit{M}${}_{2 }$(\textit{H}) component with decreasing temperature. The position of the minimum, \textit{H}${}_{min2}$, is shifted towards low fields with decreasing temperature, masking the weak-field signal from the FM clusters (see the plot at 212.2 K in Fig. 5 (a)), and reaches a minimum near 210 K, as can be seen in Fig. 6 (b). In addition, the signal amplitude at the extremum, Re \textit{M}${}_{2min2 }$(\textit{T}), exhibits a maximum at the same temperature, as follows from Fig. 6 (a). According to the arguments above, this temperature is addressed to the onset of magnetic ordering of the matrix that is \textit{T}${}_{C}$${}_{ }$$\approx$ 210 K. The extremum amplitude of the signal resulting from the FM clusters, also exhibits some enhancement near \textit{T}${}_{C}$, which is more expressed in the Re \textit{M}${}_{2}$ component. This is connected to a small but increasing contribution of the signal, coming from the matrix, to the total response. Indeed, the corresponding extremum on cooling is shifted towards weak fields, whereas its amplitude is increased. This leads to increasing contribution of the matrix signal in weak fields at \textit{T} $\rightarrow$ \textit{T}${}_{C}$ (see Fig. 6). On the other hand, such behavior gives evidence for some interaction of the FM clusters with the matrix (and for the intercluster coupling, mediated by the matrix), suggesting that a part of the FM clusters is involved into the FM ordering along with the matrix. At the same time, presence of the characteristic cluster signal well below \textit{T}${}_{C}$, following from the Re \textit{M}${}_{2}$ (\textit{H}) data at 114 K in Fig. 5 (a), indicates that some part of the clusters is not involved in the FM ordering together with matrix.

Let us discuss the behavior of the Im \textit{M}${}_{2}$   (\textit{H, T}) component of the response. It exhibits an opposite (positive) sign with respect to Re  \textit{M}${}_{2}$  (\textit{H, T}) at high temperatures, where a contribution from the matrix is not observed. This indicates that the main contribution to Im \textit{M}${}_{2}$ (\textit{H}) is connected to the effect of \textit{H}(\textit{t}) on the magnetic relaxation, given by the second term in the right-hand side of Eq. (\ref{GrindEQ__1_}). As follows from Fig. 5 (b), the signal from the matrix with a linear dependence on \textit{H} does not appear in the Im \textit{M}${}_{2}$ component on cooling even down to \textit{T} close to \textit{T}${}_{C}$. This means that the magnetic relaxation of the matrix is effective, leading to a strong inequality of $2\omega/\Gamma$ $<$$<$ 1. Therefore, according to first term in the right-hand side of Eq. (\ref{GrindEQ__1_}), the contribution of the matrix signal to Im \textit{M}${}_{2}$ ${}_{ }$$\propto$ $(2\omega/\Gamma)$ Re \textit{M}${}_{2}$ is negligible. Hence, the data of Im \textit{M}${}_{2}$(\textit{H, T}) are preferable for tracing the evolution of the cluster subsystem.

Temperature dependence of the positions, \textit{H}${}_{min1}$ and \textit{H}${}_{max}$, of the weak field extremes in the Re  \textit{M}${}_{2 }$(\textit{H}) and Im \textit{M}${}_{2 }$(\textit{H}) components, respectively, is displayed in Fig. 6 (b). It can be seen that both parameters \textit{H}${}_{min1}$ and \textit{H}${}_{max}$ are practically independent of \textit{T} both above and below \textit{T}${}_{C}$. This means that the critical properties of the FM cluster subsystem, which is responsible for the weak-field signal, are not changed within the investigated temperature interval. Such behavior is characteristic of a system of non-interacting single-domain FM particles. Taking into account a small variation of the signal amplitude in the extremum, Im \textit{M}${}_{2max}$, this behavior suggests that interaction of the FM cluster subsystem and the matrix is rather weak in the temperature range above, probably due to a specific porous structure of the samples [11]. Similar temperature behavior of the extremum positions of the \textit{M}${}_{2 }$ (\textit{H}) response from the FM clusters has been observed by us in manganites in the regime of the phase separation above \textit{T}${}_{C}$ [16, 17]. Lack of evidence for inter-cluster interaction permits us to refer the clusters to carbon particles (C-particles), which have been found in porous carbon samples doped with Ag, Au or Co, obtained by the same method as in the our case [11--13]. Indeed, according to the AMF results, C-particles observed on the sample surface are characterized by the average radius \textit{R}${}_{av}$ $\sim$ 60 nm and practically do not touch each other [11], which suggests a vanishing exchange interaction between them. Note, that the FM cluster can occupy only a part of the C-particle within its core, whereas the shell of the C-particle is characterized obviously by a larger structural disorder with respect to the core due to the lacking external bonds. This implies the different core and shell structures typical of any usual nanoparticles (see e. g. Ref. 24).

The FM clusters and the matrix make principal contributions to the signal in the weak and relatively strong fields, respectively. Therefore, it is convenient to use for analysis of the behavior of these magnetic systems the cross sections of the Re \textit{M}${}_{2}$(\textit{T, H}) data at some values of the magnetic field, \textit{H}${}_{j}$, taken below and above \textit{H}${}_{min1}$, respectively. It is interesting to obtain also the cross section of the Re \textit{M}${}_{2}$(\textit{T, H}) data at \textit{H} = 0 related to the remanent magnetization. The temperature dependences of Re \textit{M}${}_{2}$(\textit{T},  \textit{H}${}_{j}$) at \textit{H}${}_{j}$ = 0, 10 and 200 Oe are displayed for S-Au sample in Fig. 7 (b). The plot of Re  \textit{M}${}_{2}$(\textit{T}, \textit{H}${}_{j}$) vs. \textit{T} at \textit{H}${}_{j}$ = 200 Oe (where the main contribution occurs from the matrix) contains an extremum near \textit{T}$\approx$ 210 K similar to that of Re  \textit{M}${}_{2min2 }$(\textit{T}) in Fig. 6 (a). The scaling law of Eq. (\ref{GrindEQ__3_}) describes reasonably the temperature behavior above 210 K, yielding the value of \textit{T}${}_{C}$ = 209(\ref{GrindEQ__1_}) K which is in a good agreement with that found above from Re  \textit{M}${}_{2min2}$ (\textit{T}). However, the obtained value of the critical index $\gamma$${}_{2}$ = 0.7(1) is much smaller than the value, predicted for the cubic ferromagnets (14/3). This implies the percolative character of the FM ordering, connected again with a porous structure of the sample. Note, that the maximum value of Re  \textit{M}${}_{2}$(\textit{T}, 200 Oe) at \textit{T} = \textit{T}${}_{C}$ is much smaller (by $\sim${} 3 orders of the magnitude) than that of the doped cobaltites La${}_{1-x}$Sr${}_{x}$CoO${}_{3}$ (\textit{x }= 0.18 and 0.2). Both reveal a long-range percolative FM order in the ground state [25]. One order of  magnitude in above difference can be attributed to a less value of a moment of paramagnetic center in the S-Au $\sim$ 1 $\mu $${}_{B}$ (against $\sim$ 3/2 -- 2 $\mu $${}_{B}$ in manganites). However, a great residual part of this difference suggests that (i) only a small part of our sample exhibits the FM order,or the matrix and the FM cluster moments partly compensate each other e. g. due to an antiferromagnetic (AF) coupling; and/or (ii) the magnetic moment per one magnetic centre is smaller then 1 $\mu $${}_{B}$. The dependence of Re \textit{M}${}_{2}$(\textit{T, }0), which is related to the remanent magnetization, exhibits a monotonic increase with \textit{T}, whereas its value is rather close to the value of the \textit{M}${}_{2}$ response on the extremum. In addition, Figs. 6 and 7 (a) give evidence for an insignificant influence of different regimes of the temperature treatment (slow cooling or slow heating) on the \textit{M}${}_{2}$ parameters within the investigated temperature interval. This indicates a weak coupling of the magnetic moment with structural defects.

As can be seen in the inset to Fig. 6 (a), the ``coercive force'', \textit{H}${}_{C2}$, determined by the condition of Re \textit{M}${}_{2}$(\textit{H}${}_{C2}$) = 0, gradually increases with lowering the temperature. The behavior of \textit{H}${}_{C2}$ between the room temperature and \textit{T}${}_{C}$ is determined by the FM clusters, because the matrix in this temperature interval is still in the PM regime. The steep decrease of \textit{H}${}_{C2}$ near \textit{T}${}_{C}$ is connected with contribution of the non-hysteretic matrix signal, exceeding the cluster one and masking it. Below \textit{T}${}_{C}$, the onset of the domain formation should take place in the matrix. However, the development of the domain structure on cooling accompanying  usually by a pinning, occurs in our heterogeneous magnetic system in an expanded temperature region. This is evident in the plots of Fig. 5 (a) at 205.3 and 114 K, showing transformation of the hysteresis loop in the wings of the Re  \textit{M}${}_{2}$ (\textit{H}) curve on cooling. The plot at 114 K also demonstrates clear the presence of the characteristic \textit{M}${}_{2}$ signal (with extreme in a weak field) from FM clusters below \textit{T${}_{C}$}. This suggests a weak coupling of some part of the FM clusters with the matrix, as it was mentioned above. The process of domain formation in the S-Au, stretched with temperature, can be explained as follows. On one  hand, its porous structure of this sample hinders the FM ordering of the matrix in agreement with the above discussion. On the other hand,  increase of magnetostatic energy at the FM ordering stimulates to domain formatioin a sample with a single magnetic phase to decrease this energy. However, in the S-Au consisting of two magnetic subsystems, a decrease of this energy can be provided by opposite orientations of the FM matrix and the FM cluster moments. This should lead to a change of the domain formation process with respect to a sample, having a single magnetic phase. As will be evident below, such assumption is in agreement with the static magnetization data.

For the next, we discuss the \textit{M}${}_{2}$ data for sample S-Co. In Fig. 7 (b) are displayed the Re  \textit{M}${}_{2 }$(\textit{H}) signals, obtained at room temperature, from two bits (bit1 and bit2 with masses 9.8 and 14.5 mg, respectively) cut from different parts of S-Co. The signals exhibit quite different dependence on \textit{H}. In addition, the signal amplitude of bit2 is about two times higher than that of bit1 (both signals are normalized to the bit mass). This gives evidence for a large-scale magnetic (and probably structural) inhomogeneity of the sample, related presumably to a non-uniform Co ion distribution across the sample. The \textit{M}${}_{2}$ response of the bit1 exhibits a characteristic extremum in a weak field, which indicates presence of the FM clusters in this sample similar to S-Au as discussed above. This is observed on a background of the hysteretic signal from the matrix, suggesting a phase separated magnetic state of the sample formed at higher temperatures (above 350 K). The signals from the FM clusters and the matrix in bit1 of S-Co do not reveal any noticeable changes in the field dependences of both Re  \textit{M}${}_{2}$ and Im \textit{M}${}_{2}$ components of the response on cooling from 350 K down to 250 K. This implies that the magnetic states of the matrix and the cluster subsystem (as well as their mutual arrangement) are not changed within this temperature interval.

The Re \textit{M}${}_{2 }$(\textit{H, T}) response from bit2 in Fig. 8 (a) exhibits a larger magnitude and a smoother field dependence with enhanced field hysteresis, containing no typical signal from the non-interacting FM clusters. These peculiarities suggest an enhanced contribution to the response from the matrix and formation of a more homogeneous magnetic ordering, which stimulate interest for possible applications. Therefore, the data of \textit{M}${}_{2}$ for this sample are discussed below in more details. Although a signal from the FM clusters is not observed in the Re  \textit{M}${}_{2}$ component displayed in Fig. 8 (a), the presence of the FM clusters in bit2 is evident in Fig. 8 (b), which displays the Im \textit{M}${}_{2 }$(\textit{H}) signal at different temperatures. Indeed, this signal reveals the extremum in a weak field (in the interval of \textit{H} $>$ 0) again and demonstrates the opposite sign with respect to Re  \textit{M}${}_{2 }$(\textit{H}). These features are connected with the influence of \textit{H}(\textit{t}) on the relaxation rate of the magnetization of the FM cluster subsystem (the second term in Eq. (\ref{GrindEQ__1_})), similar to the Au-doped sample. No contribution to Im \textit{M}${}_{2}$(\textit{H}), which is similar to that of Re  \textit{M}${}_{2}$(\textit{H}) with respect to the sign and the dependence on \textit{H}, is observed. This indicates that the matrix contribution to Im \textit{M}${}_{2}$(\textit{H}), resulting from the delay effects, Im \textit{M}${}_{2}$(\textit{H}) $\propto$ (2$\omega$/$\Gamma$)Re  \textit{M}${}_{2}$(\textit{H}) from the first term in Eq. (\ref{GrindEQ__1_}), is small and the relaxation rate $\Gamma$ of the matrix magnetization is large. Therefore, the magnetic state of this bit is also characterized by phase separation. The noticeable features of the Im \textit{M}${}_{2 }$(\textit{H}) response from the FM cluster system are different amplitudes of the signals, recorded at direct and reverse \textit{H}-scans, as evident in Fig. 8 (b). This implies a violation of the symmetrical property of the \textit{M}${}_{2}$ response, given by the equality Im \textit{M}${}_{2 }$({--}\textit{H}) = {--} Im \textit{M}${}_{2 }$(\textit{H}). Such behavior suggests that the directions of the FM cluster moments are partially conserved, instead of their reversal at changing the direction (sign) of \textit{H}, which can be addressed to the effect of ``magnetic memory''. This phenomenon is absent in the Re  \textit{M}${}_{2}$(\textit{H}) component, because the contribution of the FM clusters to this component is negligible. The latter indicates that namely the FM clusters are responsible for such irreversibility. Possible reasons for this are attributable to a coalescence of the FM clusters (accompanied by increase of their average size), and/or increasing of their magnetic anisotropy on cooling. Similar behavior of the FM cluster response, observed in La${}_{0.78}$Ca${}_{0.22}$MnO${}_{3}$ in the vicinity of the insulator-metal transition temperature, has been attributed to the formation of a percolative network [26]. Further investigations are required to clarify the nature of the irreversibility observed in our sample.

As evident in Fig. 8 (a), the Re  \textit{M}${}_{2}$(\textit{H}) signal exhibits the hysteresis loop with only one well defined maximum at the direct \textit{H}-scan, having the amplitude Re \textit{M}${}_{2max}$ and position \textit{H}${}_{max}$ at \textit{H} $<$ 0. The Re  \textit{M}${}_{2}$ ``coercive force'',\textit{ H}${}_{C2}$, cannot be defined also in the interval of \textit{H} $>$ 0 at the direct \textit{H}-scan. The latter points out that \textit{H}${}_{C2}$ exceeds the \textit{H}-scan amplitude (i. e. \textit{H}${}_{C2 }$$>$ 300 Oe). Therefore, below we use a modified value, \textit{H${}^{*}$}${}_{C2}$, defined at \textit{H} $<$ 0. Because the FM cluster contribution to the Re  \textit{M}${}_{2}$ component is absent, the parameters \textit{H}${}_{C2}$ and \textit{H${}^{*}$}${}_{C2}$ characterize the temperature behavior of the most part of the sample, related to the matrix. As can be seen in Fig. 8 (b), the dependence of Im \textit{M}${}_{2}$(\textit{H}) at the direct \textit{H}-scan exhibits a maximum in the field interval of \textit{H} $>$ 0, Im \textit{M}${}_{2max}$, which is connected mainly to the FM clusters similar to sample S-Au. Temperature dependences of the positions and of the values of all the characteristic extremes are shown in Fig. 9. As evident in Fig. 9 (b), the positions of the extremes of Re  \textit{M}${}_{2}$(\textit{H}) and Im \textit{M}${}_{2}$(\textit{H}), found at the direct \textit{H}-scan, are practically independent of temperature. The values of the corresponding extremes exhibit a slight decrease on cooling down to $\sim$ 200 K, which is accompanied by shifting of \textit{H${}^{*}$}${}_{C2}$ towards higher field, as can be seen in Fig. 9 (a) and in the insert to Fig. 9 (b), respectively. Below 200 K, all the parameters above become almost independent of \textit{T}. This confirms that formation of the magnetic state of sample S-Co, responsible for the observed \textit{M}${}_{2}$ response, occurs appreciably above the room temperature, as has been already supposed above. Only a minor modification of this state (connected to the increase of pinning) is observed on cooling down to 200 K, where it is stabilized and does not change at least down to 100 K. The temperature behavior and the value of Re  \textit{M}${}_{2}$(\textit{H}=0) are quite similar to those of Re  \textit{M}${}_{2max}$(\textit{T}), which indicates a large value of the remanent magnetization. Note, the temperature evolution of the \textit{M}${}_{2}$ response in bit1 of sample S-Co in the interval of 350 K $\geq$ \textit{T} $\geq$ 250 K is similar to that of bit2, exhibiting only a minor signal variation without any qualitative transformation. This suggests that the magnetic state of both bits does not change on cooling within the investigated temperature interval.

Comparison of the \textit{M}${}_{2}$ data for the samples doped by Au and Co, respectively, demonstrates that the amplitude of the \textit{M}${}_{2}$ response in S-Co is enhanced by $\sim$ 5 times. Such enhancement is attributable to the larger sizes of the structural clusters, formed in sample S-Co at synthesis, which is supported by the neutron diffraction data. A difference of the amplitude of the \textit{M}${}_{2}$ response in the Co-doped sample and in typical cobaltite La${}_{0.8}$Sr${}_{0.2}$CoO${}_{3}$ with the FM ground state [24] is decreased with respect to the Au-doped sample, but still consists of $\sim$ 2 orders of the magnitude. 

\subsection{ Electron magnetic resonance }

In Fig. 10 are displayed the EMR spectra of the samples S-Au and S-Co (bit2) at room temperature. As evident in Fig. 12 (a), the Au-doped sample exhibits a single narrow line with the Land\'e factor \textit{g} = 2.0324(8). This indicates the presence of PM centers not involved in formation of the FM clusters. These centers can be attributed to the PM matrix in agreement with the \textit{M}${}_{2}$ data. The absence of signals from the FM clusters is connected probably with their insufficient density. Another possible reason is the large transversal magnetic relaxation of cluster subsystem, related to a wide space distribution of the magnetic anisotropy axes and to the fluctuating values of the anisotropy constants in different parts of the sample with amorphous structure. The S-Au sample exhibits also a contribution to the EMR spectrum with a linear dependence on the external steady field, \textit{H}. The origin of the linear contribution is connected with the Hall effect on microwave frequency due to conduction electrons. The Hall signal on microwave frequency is also detected by our home made spectrometer, since it is provided by a cylindrical balanced cavity with H${}_{111}$ mode of the electromagnetic oscillations. This possibility was confirmed earlier by registration of the Hall signal from a material (Cu) of the cavity [18]. The presence of such a signal from the S-Au sample suggests its electrical conductivity. 

In Fig.10 (b) the EMR spectrum is displayed from bit2 of the Co-doped sample at room temperature. The spectrum contains a single wide line, which can be interpreted as a signal of a FM resonance from this sample. The signal suggests an almost homogeneous FM ordering of bit2, which have been established above 300 K. Hence, the spectrum in Fig. 10 (b) confirms the conclusion about the magnetic state of the sample S-Co, made above from the analysis of the \textit{M}${}_{2}$ data.

\subsection{Magnetization measurements.}

Temperature dependence of the \textit{dc} magnetization in the Au-doped sample, measured in the fields of $\sim$ 50 mT and 1 T, is characterized by the deviation of \textit{M}${}_{ZFC}$ (\textit{T}) from \textit{M}${}_{FC}$ (\textit{T}), and by TRM (see Fig. 11 (a)). The absolute value of the difference between \textit{M}${}_{ZFC}$ (\textit{T}) and \textit{M}${}_{FC}$ (\textit{T}) initially grows with increasing field up to a maximum at \textit{B }$\sim$ 0.2 T, then it decreases and vanishes at \textit{B }above $\sim$ 1 T. The relative difference of \textit{M}${}_{ZFC}$ (\textit{T}) and \textit{M}${}_{FC}$ (\textit{T}) with respect to \textit{M}${}_{FC}$ decreases monotonically with increasing \textit{B}, revealing a faster variation below a field of $\sim$ 0.2 T. TRM is shifted gradually towards higher values when \textit{B} is increased up to 0.2 T. The dependences of TRM(\textit{T}), obtained between \textit{B} $\sim$ 0.2 and 1 T, practically coincide with each other (see Fig. 12 (a)), whereas \textit{M}${}_{ZFC}$(\textit{T}) and \textit{M}${}_{FC}$(\textit{T}) persist to increase. This means that after switching off the field, the system tends to equilibrium determined only by temperature. Below \textit{B} $\sim$ 0.2 T, a state of the sample after the field switching off depends on \textit{T}, as well as on the field value. The latter indicates some quasi-equilibrium state of the magnetic system. The magnitude of TRM exhibits a monotonic increase with decreasing \textit{T} between 300 and 140 K and a crossover to a faster upturn below \textit{T}${}_{cr}$ $\approx$ 140 K. Eventually, a steep increase of TRM (\textit{T}) is observed below \textit{T} $\sim$ 3 -- 10 K and \textit{B} between $\sim$ 0.2 {--} 1 T. The dependence of \textit{M}${}_{ZFC}$(\textit{T}), measured in the low field of \textit{B} = 46.5 mT, does not reveal the increase below \textit{T}${}_{C}$ expected during domain formation. In turn, the value of \textit{M}${}_{ZFC}$ at \textit{T} $\sim$ 140 K, coinciding with the crossover temperature, \textit{T}${}_{cr}$, of TRM (\textit{T}), exhibits a rather steep decrease in all fields below \textit{B} $\sim$ 0.5 T. These features reflect an unusual character of the magnetic ordering in the S-Au sample. A possible reason to such an ``antiferromagnetic'' (AF) behavior of \textit{M}${}_{ZFC}$(\textit{T}) can be addressed to a peculiar AF interaction of the ferromagnetically ordered matrix and the FM cluster subsystem, which is attributable to the C-particles (see Subsection 3). The AF interaction of the C-particle FM core and an adjoining part of the matrix may be provided by the C-particle shell. The  latter has probably a different structure and magnetic properties like that in usual magnetic nanoparticles [24]. 

The AF interaction induces above \textit{T}${}_{C}$ AF correlations in the matrix regions near the C-particles, which compete with the FM correlations opposing their fast growth. This leads to reduction of the critical indexes of the PM-FM transition in the matrix, which is in agreement with a small value of $\gamma$${}_{2}$ following from the analysis of the \textit{M}${}_{2 }$ data. The AF correlations persist to increase on cooling below \textit{T}${}_{C}$ leading to a peculiar ``AF ordering'' of the matrix and the FM cluster moments below the crossover temperature \textit{T}${}_{cr}$ $\sim$ 140 K. This explains the decrease of \textit{M}${}_{ZFC}$, observed at \textit{B} = 46.5 mT below \textit{T}${}_{cr}$, as shown in Fig. 11(a). Below \textit{T}${}_{C}$, S-Au sample can be considered as a ``\textit{peculiar ferrimagmet}'', containing two unusual magnetic sublattices. A state of the first sublattice (related to the matrix) is close to a complete FM ordered, whereas the second sublattice consists of the randomly distributed and very weakly interacted FM clusters. Slightly below \textit{T}${}_{C}$, the matrix is built probably from ferromagnetically ordered fragments, surrounded by the FM clusters with moments oriented mainly opposite to that of the matrix fragments due to their AF correlations. Such a composite structure can be considered as a peculiar ``domain'', since its formation is accompanied by a decrease of the magnetostatic energy. The AF ordering below 140 K leads to an almost collinear alignment of the cluster moments and the FM moment of a matrix fragment in each ``domain''. This takes place along with weakening a weak FM coupling between such ``domains'' in porous amorphous structure. Surely, a complete compensation of the corresponding FM moments in a ``domain'' does not occur. Therefore, a weak moment still exists, which is an uncompensated remainder either of the moment of a matrix fragment, or of a total moment of FM clusters, surrounding this fragment. The resulting moments of the ``domains'' can have opposite signs, since a number of surrounding clusters, as well as a size of the matrix fragment and its moment can differ. Besides, their uncompensated moments can be oriented almost randomly due to a weak coupling of the ``domains'' in a porous amorphous structure of the S-Au sample. The proposed AF ``ordering'' leads to a decrease of \textit{M}${}_{ZFC}$ with cooling below 140 K. Note, that only a part of the FM clusters are involved in formation of the ``domains''. This follows from the Re (Im  )\textit{M}${}_{2}$(\textit{H}) dependences, which exhibit a presence of the characteristic signal with extremum in a weak field, pertinent to FM clusters at \textit{T} = 114 K below \textit{T}${}_{cr}$ (Fig. 5). The formation of the ``domains'' is accompanied by a pinning, which appears below 140 K along with increasing TRM Fig. 12(a), and the field hysteresis of the \textit{M}${}_{2}$ response, as evident in Fig. 5(a) at \textit{T} = 114 K. Such a peculiar AF ``ordering'' explains, at least partly, a small magnitude of the \textit{M}${}_{2}$ signal connected with matrix (see Subsection 3). On cooling, even a weak external field of \textit{B} = 46.5 mT  makes such mechanism less effective, as can be seen in Fig. 11(a). In the FC regime, the external field provides a partial alignment of the moments of FM clusters along the field above \textit{T}${}_{C}$, as well as an alignment of the uncompensated moments of the ``domains'' below \textit{T}${}_{C}$. This explains the predominance of \textit{M}${}_{FC}$ over \textit{M}${}_{ZFC}$. Increase of the external field leads to the corresponding increase of M${}_{FC}$(\textit{T}) due to better alignment of  the moments of FM clusters and of the ``domains''. On the other hand, application of the same increased field for measurements of \textit{M}${}_{ZFC}$(\textit{T}) hinders a pinning of the randomly oriented domains below \textit{T}${}_{C}$. Above \textit{T}${}_{C}$ this leads to a better orientation of the FM moments of C-particles along the field. Therefore, the increasing \textit{B} decreases the difference of \textit{M}${}_{ZFC}$(\textit{T}) and \textit{M}${}_{FC}$(\textit{T}) at any \textit{T}. Note, that the anisotropy of any magnetic nanoparticles is attributed to a type of an easy axis [23].In a zero external field the magnetic moment of a nanoparticle is directed along this axis providing a divergence between \textit{M}${}_{ZFC}$(\textit{T}) and \textit{M}${}_{FC}$(\textit{T}) at temperatures below blocking temperature \textit{T}${}_{b}$ [23]. Their coincidence is achieved at \textit{T} $>$ \textit{T}${}_{b}$ and our data permits us to estimate the mean anisotropy field, \textit{B}${}_{a}$, of the C-particles to satisfy the relation of   0.2 $<$ \textit{B}${}_{a}$ $<$ 0.5 T. Indeed, relative difference of \textit{M}${}_{ZFC}$(\textit{T}) and \textit{M}${}_{FC}$(\textit{T}) below the external field of $B$ = 0.2 T is practically constant, whereas at \textit{B} $>$ 0.2 T it decreases steeply both above and below \textit{T}${}_{C}$. On the other hand, the difference of \textit{M}${}_{ZFC}$(\textit{T}) and \textit{M}${}_{FC}$(\textit{T}) at \textit{B} = 0.5 T is close to zero above \textit{T}${}_{C}$. In addition, the coincidence of \textit{M}${}_{ZFC}$(\textit{T}) and \textit{M}${}_{FC}$(\textit{T}) below \textit{T}${}_{cr}$, which is evident in Fig. 11(a) at \textit{B} $\sim$ 1 T near \textit{T} $\sim$ 20 K permits us to estimate a mean effective field of the pinning in the interval of  \textit{B} $\sim$ 0.2 T -- 1 T. 

On cooling down to \textit{T} $\sim$ 3 K, \textit{M}${}_{ZFC}$(\textit{T}) and \textit{M}${}_{FC}$(\textit{T}) in Fig. 11 (a) begin to increase, indicating an onset of the magnetic rearrangement in the S-Au. The latter can be interpreted as a transition from an almost opposite orientation of the matrix and the FM cluster moments in the ``domains'' to an nearly parallel alignment. This takes place along with formation of a new domain system, which looks more similar to that of an ordinary ferromagnet. Evidently, such arrangement is more preferable for minimization of the free energy of the sample at temperature \textit{T}${}_{MT}$ $\sim$ 3 K, looking like a transition from the ferrimagnetic to the FM state in a compound with two FM subsystems having different magnetic moments [27]. This transition depends on the applied magnetic field, and its onset shifts towards a higher temperature with increasing \textit{B}, which suggests again a quasi-equilibrium magnetic state of the sample.

The unusual magnetic state of the S-Au sample is indicated also by the change of sign of the field hysteresis, which is observed in the dependence of\textit{ M}(\textit{B}) at \textit{B} $\sim$ 1 T and \textit{T} = 5.1 K $>$ \textit{T}${}_{MT}$, as can be seen in the insert to Fig. 11 (b). Here, the difference between \textit{M}${}_{ZFC}$(\textit{T}) and \textit{M}${}_{FC}$(\textit{T}) becomes negligible, as follows from Fig. 11 (a). Several \textit{M }(\textit{B}) curves, measured at temperatures from different characteristic intervals of \textit{T $>$ T}${}_{C}$, \textit{T}${}_{C }$$>$ \textit{T }$>$ \textit{T}${}_{cr}$, \textit{T}${}_{cr }$$>$ \textit{T }$>$ \textit{T}${}_{MT}$${}_{ }$ and at \textit{T} $\sim$ \textit{T}${}_{MT}$, are displayed in Fig. 11 (b). A steep increase of \textit{M}(\textit{B}) up to approximately the same value at \textit{T }= 225 K \textit{$>$} \textit{T}${}_{C}$, \textit{T}${}_{C }$$>$ \textit{T }= 180 K $>$ \textit{T}${}_{cr}$, and even at \textit{T} = 100 K $<$ \textit{T}${}_{cr }$in low fields of \textit{B }$\leq$ 0.8 T is observed. This implies that the main contribution to the magnetization in the indicated field interval is given by the individual FM nanoparticles with different sizes, which are not involved in the magnetic arrangement discussed above. Close values of \textit{M}(\textit{T}) at any \textit{B }$\geq$ 1 T for curves obtained at \textit{T} = 225 and 180 K $>$ \textit{T}${}_{cr }$, give evidence for a presence of the AF correlations above \textit{T}${}_{cr}$. At lower \textit{T} = 100, 50 and 5 K below \textit{T}${}_{cr}$, a crossover to a moderate and approximately linear increase of \textit{M} (\textit{B}) in the interval of \textit{B }$\geq$ 1 T is observed up to 5 T. It is worth mentioning, that a similar dependence of \textit{M} (\textit{B}) with a steep increase in a weak field, accompanied by the crossover to a linear behavior up to a high \textit{B} $\sim$ 10 T, is observed in ferrites with a canted configuration of sublattices [27]. Such a configuration is quite expectable for the peculiar sublattices in our porous S-Au sample, confirming our model. Cooling the simple between 100 and 5 K leads to a shift of \textit{M} (\textit{B}) to higher values, which is connected probably with a continuation of the ``domain'' formation.

On further cooling down to the onset  of the magnetic transition to FM alignment of the matrix and the FM cluster subsystem moments in the S-Au sample (at \textit{T $\sim$ T}${}_{MT}$ $\sim$ 3 K), a considerable growth of \textit{M} (\textit{B}) is observed. However, \textit{M} (\textit{B}) is still far from saturation at the highest \textit{B} $\sim$ 5 T, which is evident in Fig. 11 (b). As follows from the dependences of \textit{M}${}_{ZFC }$ (\textit{T}) and \textit{M}${}_{FC }$ (\textit{T}) at \textit{B} = 46.5 mT displayed in Fig. 11 (a), the rearrangement of the magnetic ordering in the S-Au sample only begins at \textit{T}${}_{MT}$ $\sim$ 3 K. This suggests that magnetic organization of any ``domain'' differs from that described above only insignificantly. Namely this can lead to the absence of saturation of the magnetization in high fields. At the same time, the onset of the transition leads to a considerable enhancement of \textit{M} (\textit{T}) and more smooth (without a crossover) behavior of \textit{M}(\textit{B}). The latter can be seen in Fig. 11 (b)  at 3.3 K. Hence, the \textit{M} (\textit{B}) data support our interpretation as well. The presence of the AF correlations above \textit{T}${}_{C}$ and formation of the AF ``domains'' below \textit{T}${}_{C }$ can give an explanation of the small amplitude of the \textit{M}${}_{2}$ response in the S-Au sample (see Subsection 3).

Similar peculiarities can be seen also in the dependences of \textit{M}${}_{ZFC}$ (\textit{T}), \textit{M}${}_{FC}$ (\textit{T}) and TRM (\textit{T}) of the undoped powder carbon-based sample [11], suggesting a magnetic state similar to our S-Au sample and a weak effect of Au doping. In the Co-doped sample, the crossover of \textit{M}${}_{ZFC}$(\textit{T}) and TRM (\textit{T}) is not found below 350 K. However, there exist such features as the divergence of \textit{M}${}_{ZFC}$(\textit{T}) and\textit{M}${}_{FC}$(\textit{T}), the sensitivity of these parameters and of TRM(\textit{T}) to the applied magnetic field.  Finally, the steep increase of \textit{M}${}_{ZFC}$(\textit{T}) and \textit{M}${}_{FC}$(\textit{T}) on cooling down to \textit{T} $\sim$ 10 K, depending on \textit{B}, are observed in the S-Co sample, as can be seen in Fig. 12 (b). These results suggest formation of the specific heterogeneous magnetic state of the S-Co sample, which is similar to that of the S-Au sample, at substantially higher temperatures in agreement with the \textit{M}${}_{2}$ data discussed above. This assumption permits us to explain a small value of the \textit{M}${}_{2}$ response in the S-Co sample in comparison with doped cobaltites (see Subsection 3), by formation of the AF ``domains'' similar to the S-Au sample. The latter occurs, however, at higher \textit{T} exceeding 350 K in agreement with the absence of the crossover to a faster decrease in \textit{M}${}_{ZFC}$(\textit{T}), as can be seen in Fig. 12 (b).

Comparison of the TRM(\textit{T}) data at \textit{B} = 1 kG in Figs.12 (a) and (b), as well as of the \textit{M}${}_{FC}$(\textit{T}) and \textit{M}${}_{ZFC}$(\textit{T}) data in similar fields of \textit{B} = 0.5 and 1 kG and at close temperatures demonstrate, that the values of all these parameters in the Co-doped sample are much smaller than in the Au-doped one. This is in disagreement with the results of the nonlinear response discussed above, as well as with the magnetization data for similar samples in [11--13], where the corresponding relations are inverted. These contradictions are attributable to the large-scale spatial inhomogeneity of the Co-doped sample, following from the \textit{M}${}_{2}$ data discussed above, since our magnetization measurements have been carried out using another bit of this sample. Indeed, comparison of the \textit{M}${}_{2}$ data obtained in different parts of the sample S-Co (bit1 and bit2) demonstrates (see Subsection 3), that the temperature evolution of their magnetic states on cooling is similar and the relative difference of the signal amplitudes is conserved at high temperatures. In addition, the observed behavior of the magnetization of the Co-doped sample on cooling is similar to that observed in [11-13]. These results suggest that the disagreements above do not change qualitatively the scenario of the temperature evolution of magnetic state in different parts of the Co-doped sample. Therefore, we have used the magnetization data of the S-Co sample, obtained here, in the comparative qualitative analysis above. 

\section{Conclusions}

According to the neutron diffraction data, the structure of both nanocarbon samples, investigated in this work, has the amorphous character. This corresponds to the well-known concept of organization of the carbon-metal nanocomposites, containing nanoporosity. The Co-doped sample exhibits a more regular distribution of pores and probably larger average sizes of the sample material outside the pores with respect to the sample doped with Au. This is accompanied by a more homogeneous short-range magnetic arrangement, as well as by formation of a ground magnetic state with a long-range FM ordering, which is well detected by the neutron diffraction study.

NMR investigations of the local structure of the samples permit us to conclude, that they are (i) the products of partial carbonization of initial aromatic compounds and (ii) these products have not reach a state of glassy carbon. 

The main result of the magnetic investigations of   composite samples doped with Au and Co is establishing of their inhomogeneous phase-separated magnetic state, which depends on temperature. This state contains the system of the FM cluster and the magnetic matrix. The latter is formed by paramagnetic centers located outside the FM clusters. The magnetic characteristics and their temperature behavior, as well as structure of the compounds depend appreciably on the doping material. In the sample doped by nonmagnetic Au, the onset of the matrix ordering occurs at lower temperature (\textit{T}${}_{C}$ $\approx$ 210 K) whereas in the Co-doped sample this ordering takes place at higher temperature above 350 K. The S-Co sample exhibits the remanent magnetization and the coercive force, which exceed considerably those of the S-Au sample. In addition, the Co-doped sample displays inhomogeneous magnetic properties on the long-range spatial scale, characterized by larger magnitude of the mean magnetic moment. The complex temperature behavior of the magnetization in the Au-doped sample suggests changing of a mutual arrangement of magnetic moments of the matrix and the FM cluster system from an almost opposite orientation below \textit{T}${}_{C}$ to an almost parallel one at low temperatures. Only the last stage of this process has been observed in the S-Co sample within the investigated temperature interval. This stage is accompanied probably by formation of an almost homogeneous FM state, as follows from the neutron diffraction investigations.

Generally, the results obtained by different techniques permit us to clarify the peculiarities of the structure and to obtain important information about delicate processes of the magnetic arrangement of carbon-based porous nanomaterials doped by Au and Co.

\noindent \textbf{}

\noindent \textbf{Acknowledgments}

\noindent \textbf{}

\noindent The work was carried out under partial support of the St. Petersburg State University and the Program of the Presidium RAS ``Bases of fundamental studies of nanotechnologies and nanomaterials'' (Project No. 4.4.1.8).

\noindent \textbf{}

\noindent \textbf{References }

\noindent \textbf{}

\noindent [1]  T. Makarova and F. Palacio (eds.). \textit{Carbon-Based Magnetism} (Elsevier, North-Holland,  2006).

\noindent [2]  T. Makarova. \textit{Unconventional magnetism in carbon based materials, in: Frontiers in Magnetic Materials}. A.V. Narlikar (ed.), p. 5, (Springer, Berlin, 2005).

\noindent [3]  N. Park, M. Yoon, S. Berber, J. Ihm, E. Osawa and D. Tomanek. Phys. Rev. Lett. \textbf{91}  237204 (2003).

\noindent [4]  A. A. Ovchinnikov and V. N. Spektor. Synth. Mather. \textbf{27}  B615 (1988).

\noindent [5]  Y.-W. Son, M. L. Cohen and S. G. Louie. Nature
\textbf{444}  347 (2006).

\noindent [6]  O. V. Yazyev and M. I. Katsnelson. Phys. Rev. Lett. \textbf{100}  047209 (2008).

\noindent [7]  O. V. Yazyev. Phys. Rev. Lett. \textbf{101}  37203 (2008).

\noindent [8]  R. Faccio, H. Pardo, P. A. Denis, R. Y. Oeiras, F. M. Araujo-Moreira, M. Verissimo-Alves and A. W. Mombru. Phys. Rev. B \textbf {77}  035416 (2008).

\noindent [9]  O. V. Yazyev and L. Helm. Phys. Rev. B \textbf{75}  125408 (2007).

\noindent [10] J. \v{C}ervenka, M. I. Katsnelson and C. F. J. Flipse. Nature Physics \textbf{5}  840 (2009). 

\noindent [11] E. L\"ahderanta, A. V. Lashkul, K. G. Lisunov, A. Pulkkinen, D. A. Zherebtsov, D. M. Galimovand A. N. Titkov. IOP Conf. Series: Materials Science and Engineering \textbf{38} ( 012010 2012).

\noindent [12] E. L\"ahderanta, A. V. Lashkul, K. G. Lisunov, D. A. Zherebtsov, D. M. Galimov, A. N. Titkov, EPJ Web of Conferences \textbf{40}  08008 (2013).

\noindent [13] E. L\"ahderanta, A. V. Lashkul, K. G. Lisunov, D. A. Zherebtsov, D. M. Galimov, A. N. Titkov, J. Nanosci. Nanotech. \textbf{12}  9156 (2012).

\noindent [14] D. Massiot, F. Fayon, M. Capron, I. King, S. Le Calv\'e, B. Alonso, J. O. Durand, B. Bujoli, Z. Gan, G. Hoatson. Modelling one and two-dimensional solid-state NMR spectra. Magn. Reson. Chem. \textbf{40}  70--76 (2002).

\noindent [15] V.A. Ryzhov, I.I. Larionov, V.N. Fomichev. Zh. Tekh. Fiz. \textbf{66 } 183 (1996) [Sov.Phys. Tech. Phys. \textbf{41}  620 (1996)]; V.A. Ryzhov, E. I. Zavatskii. Patent No 2507525, registered 20.02.2014.

\noindent [16] V.A. Ryzhov, A.V. Lazuta, I.D. Luzyanin \textit{et al}. JETP \textbf{94} (2002) 581-592; V.A. Ryzhov, A.V. Lazuta, V.P. Khavronin \textit{et al}. Sol. St. Comm.\textbf{130}  803--808 (2004). 

\noindent [17] V.A. Ryzhov, A.V. Lazuta, V.P. Khavronin \textit{et al}. JMMM\textbf{300}  e159--e162 (2006).

\noindent [18] V. A. Ryzhov, E. I. Zavatskii, V. A. Solov'ev, I. A. Kiselev, V. N., Fomichev, and V. A. Bikineev. Zh. Tekh. Fiz. \textbf{65}  133 (1995) [Tech. Phys. \textbf{40}  71 (1995)].

\noindent [19] J. Herzfeld and A. E. Berger. Sideband Intensities in NMR Spectra of Samples Spinning at the Magic Angle. J. Chem. Phys.\textbf{73},  6021-6030 (1980). 

\noindent [20] Hellmut Eckert, Yiannis A. Levendis, Richard C. Flagan. J. Phys. Chem. \textbf{92}  5011--5019 (1988). 

\noindent [21] M. V. Mokeev, A. V. Gribanov, Yu. N. Sazanov. \textit{Russian Journal of Applied Chemistry} \textbf{84} (2011) 111--117.

\noindent [22] A. V. Lazuta, I. I. Larionov, and V. A. Ryzhov. Zh. Eksp. Teor. Fiz. \textbf{100}  1964 (1991) [Sov. Phys. JETP \textbf{73}  1086 (1991)].

\noindent [23] B. D. Gullity, C. D. Graham. \textit{Introduction to Magnetic Materials, Second Edition.} (Wiley John \& Sons, Hoboken, IEEE Press, New Jersey,  2009).

\noindent [24] I.V. Golosovsky, G. Salazar-Alvarez, A. L\'opez-Ortega,M. A. Gonz\'alez,J. Sort,M. Estrader,S. Suri\~nach,M. D. Bar\'o,and J. Nogu\'es. Phys. Rev. Lett. \textbf{102}, 247201 (2009).

\noindent [25] A.V. Lazuta, V.A. Ryzhov, A.I. Kurbakov, V.P. Khavronin, P.L. Molkanov,Ya.M. Mukovskii, A.E. Pestun and R.V. Privesentsev. Solid State Phenomena \textbf{168-169}  457--460 (2011).

\noindent [26] V.A. Ryzhov, A.V. Lazuta, O.P. Smirnov, V.P. Khavronin, P.L. Molkanov, Ya.M. Mukovskii and V.I. Chichkov. Solid State Phenomena \textbf{168-169}  485--488 (2011). 

\noindent [27] S. V. Vonsovskii.\textit{ Magnetism.} (Moskow, ``Nauka'', 1971).

\noindent \textbf{}

\noindent \textbf{ Table 1. Composition of the S-Au sample}  \\

\begin{tabular}{|p{1.1in}|p{1.1in}|p{1.1in}|} \hline 
Line position\newline  (ppm) & Type of fragment & Relative content\newline (\%) \\ \hline 
33 & aliphatic fragments & 2 \\ \hline 
126 & aromatic fragments & 81 \\ \hline 
136 & Ar-O, Ar-C,=C- & 12 \\ \hline 
152 & $>$C=O, $>$C=N- & 5 \\ \hline 
\end{tabular}
 \\
 \\
 
\noindent \textbf{Figure captions}

\noindent \textbf{}

\noindent \textbf{Figure 1.} The neutron diffraction patterns, obtained at 300 K for the S-Au sample (curve 1), and at 300 K (curve 2) and 2.6 K (curve 3) for the S-Co sample. They are normalized on the same time of measurements. The curve 4 is the difference of the curves 3 and 2. 

\noindent \textbf{Figure 2.} ${}^{1}$H MAS NMR spectrum of the sample S-Au. Two main lines are at 2.3 ppm and 6.8 ppm (see the text). Other lines correspond to artifacts due to the spinning (Spinning Side Bands, SSB). 

\noindent\textbf{Figure 3.} The ${}^{13}$C CPMAS NMR spectrum of the sample S-Au.

\noindent \textbf{Figure 4.} The ${}^{13}$C NMR spectrum of the sample S-Au obtained using the direct-acquisition method with proton decoupling (\textit{hpdec)}. The 111 ppm peak corresponds to the signal of the probe head.

\noindent \textbf{Figure 5.} The dependences of Re \textit{M}${}_{2}$ and Im \textit{M}${}_{2 }$ on the steady magnetic field \textit{H} at different temperatures for the Au-doped sample with \textit{m} = 9.8 mg. The closed and the open symbols represent the data recorded at the direct and the reverse \textit{H}-scans, respectively.

\noindent \textbf{Figure 6.} The extremum values of Re \textit{M}${}_{2}$ (\textit{H}) and Im \textit{M}${}_{2}$ (\textit{H}) (a) and the corresponding positions of the extremes (b) as functions of the temperature for the Au-doped sample (S-Au). Insert to the panel (a): The temperature dependence of the ``coercive force'', \textit{H}${}_{C2}$, of Re \textit{M}${}_{2}$ (\textit{H}). The parameter \textit{H}${}_{C2 }$ is determined by the condition of Re  \textit{M}${}_{2}$ (\textit{H}${}_{C2}$) = 0.

\noindent \textbf{Figure 7.} Panel (a): The dependence of Re \textit{M}${}_{2 }$ (\textit{T},\textit{ H}${}_{j}$) on temperature at \textit{H}${}_{j}$ = 0, 10 and 200 Oe. The solid line is the fit of Re \textit{M}${}_{2}$ (\textit{T}, 200 Oe) with a scaling law, \textit{M}${}_{02}$ \textit{$\tau$${}^{-}$${}^{\gamma}$}${}_{2}$, for the S-Au sample. Panel (b): The Re \textit{M}${}_{2}$ (\textit{H}) response (the direct and the reverse \textit{H}-scans are given by the close and the open symbols, respectively) of two bits of the Co-doped sample at room temperature. The parameters of the Re \textit{M}${}_{2}$ (\textit{H}) signal, which are usually used for its characterization, are exhibited. The similar parameters are used for characterization of the Im \textit{M}${}_{2}$ (\textit{H}) component, as well.

\noindent \textbf{Figure 8.} The dependences of Re   \textit{M}${}_{2}$ and Im \textit{M}${}_{2}$ on the steady magnetic field \textit{H} for the Co-doped sample at different temperatures. Closed and open symbols are used for curves recorded at direct and reverse \textit{H}-scans, respectively.

\noindent \textbf{Figure 9.} Temperature dependences of the extremum values (at direct \textit{H}-scan) Im \textit{M}${}_{2max}$, Im \textit{M}${}_{2min}$ and Re  \textit{M}${}_{2max}$, and of Re  \textit{M}${}_{2}$ (\textit{H} = 0) in the Co-doped sample (a). Temperature dependences of the positions of these extremes, \textit{H}${}_{max }$(of Re  \textit{M}${}_{2}$ and Im \textit{M}${}_{2}$) and \textit{H}${}_{min}$ (of Im \textit{M}${}_{2}$) (b). Insert in panel (b) displays the dependence of \textit{H}${}_{C2 }$ in Re \textit{M}${}_{2}$ on \textit{T} for this sample. The data obtained in the regimes of slow cooling and slow heating are shown.

\noindent \textbf{Figure 10.} Room temperature EMR spectra of the S-Au sample with m = 8 mg (a) and of the S-Co sample with m = 14.5 mg (b).

\noindent \textbf{Figure 11.} Temperature dependences of \textit{M}${}_{ZFC}$, \textit{M}${}_{FC}$ and TRM for the S-Au sample . The plots at 1 T are multiplied by a factor of 0.3 for convenience (a). The dependence of \textit{M }on \textit{B} for the S-Au sample at different temperatures. Inset: The hysteresis loop of \textit{M }(\textit{B}) at \textit{T} = 5.1 K (b).

\noindent \textbf{Figure 12.} The plots of TRM vs. \textit{T }in different fields for the S-Au sample (a). The temperature dependences of \textit{M}${}_{ZFC}$, \textit{M}${}_{FC}$ and TRM for the S-Co sample (b).
\begin{center}
\includegraphics*[width=1.3\linewidth, keepaspectratio=true]{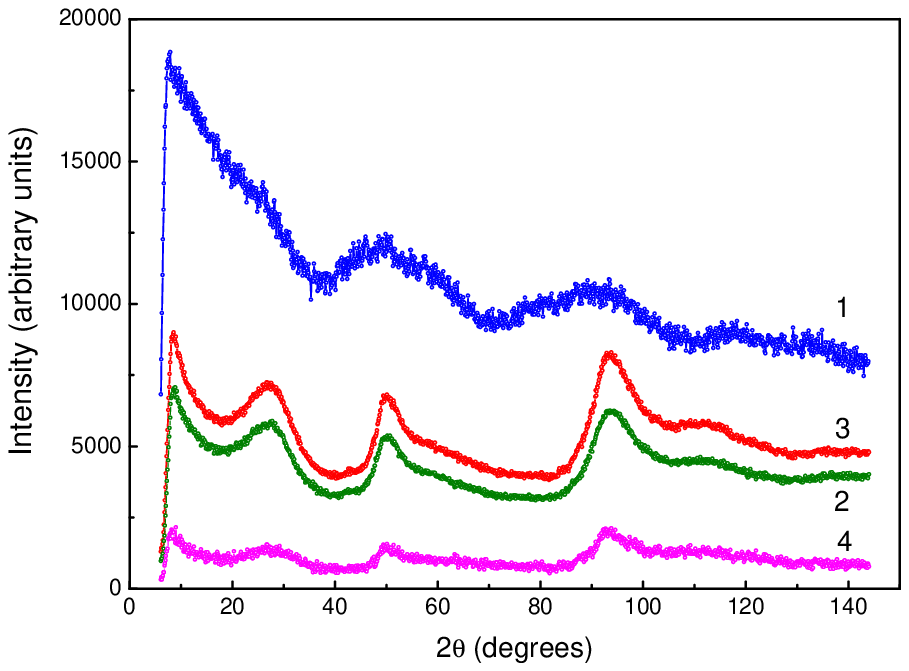}
\textbf{Figure 1.}
\end{center}

\begin{center}
\includegraphics*[width=1.3\linewidth, keepaspectratio=true]{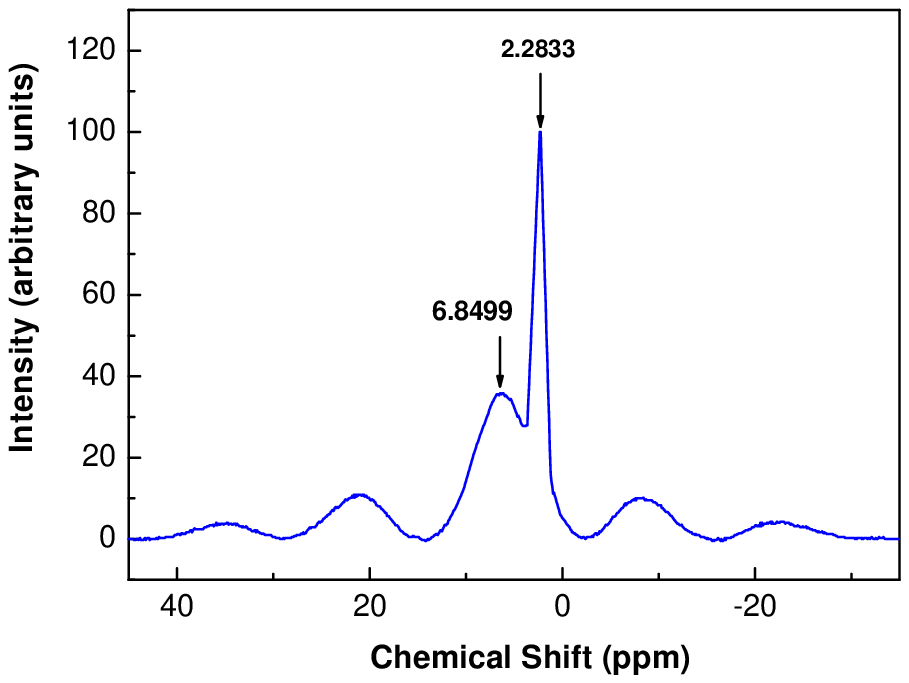}
\textbf{Figure 2.}
\end{center}

\begin{center}
\includegraphics*[width=1.3\linewidth, keepaspectratio=false]{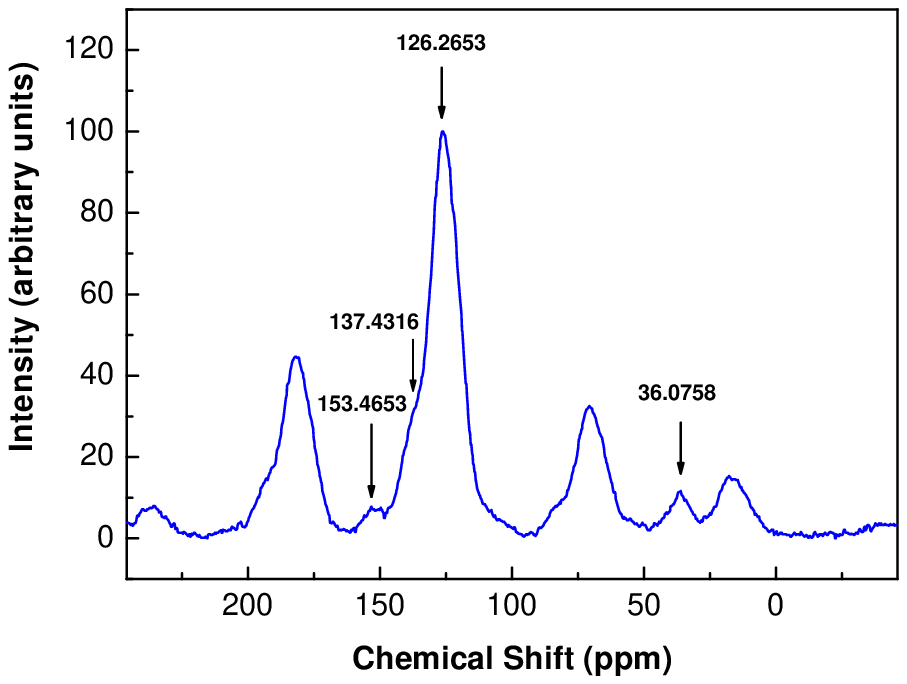}
\textbf{Figure 3.}
\end{center}

\begin{center}
\includegraphics*[width=1.3\linewidth, keepaspectratio=false]{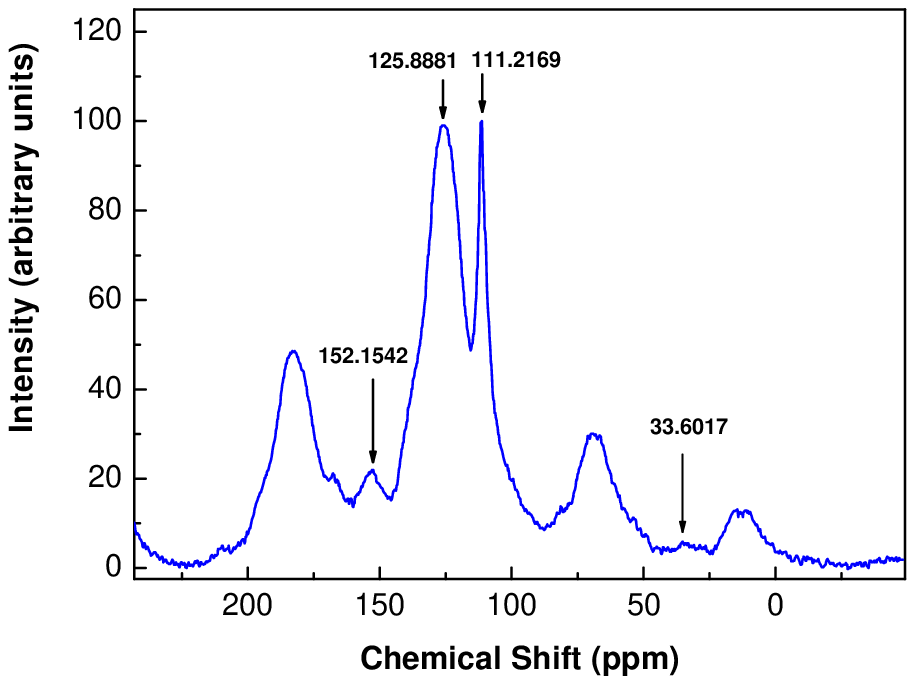}
\textbf{Figure 4.}
\end{center}

\begin{center}
\includegraphics*[width=1.3\linewidth, keepaspectratio=false]{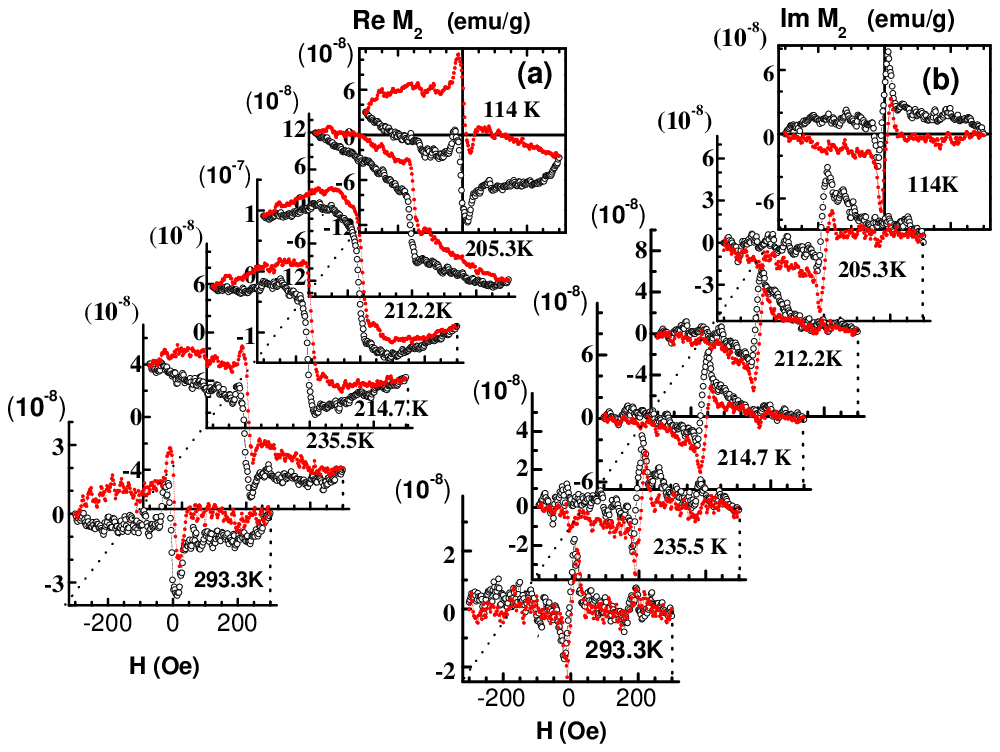}
\textbf{Figure 5.}
\end{center}

\begin{center}
\includegraphics*[width=1.3\linewidth, keepaspectratio=false]{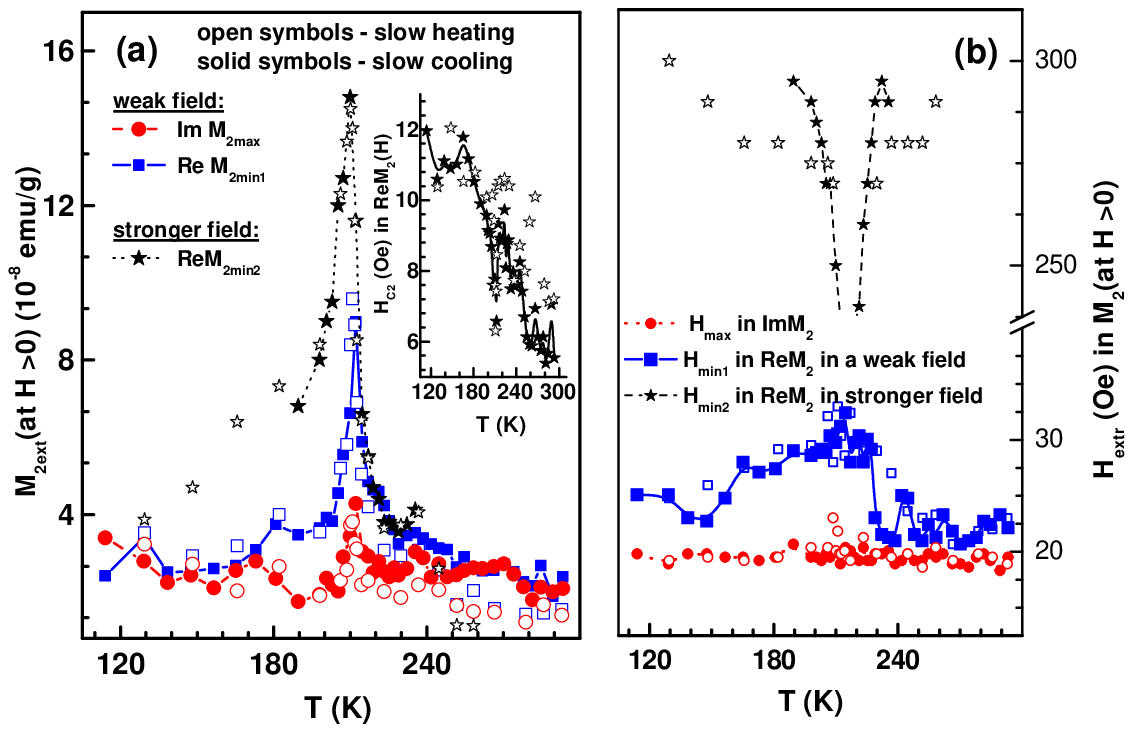}
\textbf{Figure 6.}
\end{center}

\begin{center}
\includegraphics*[width=1.3\linewidth, keepaspectratio=false]{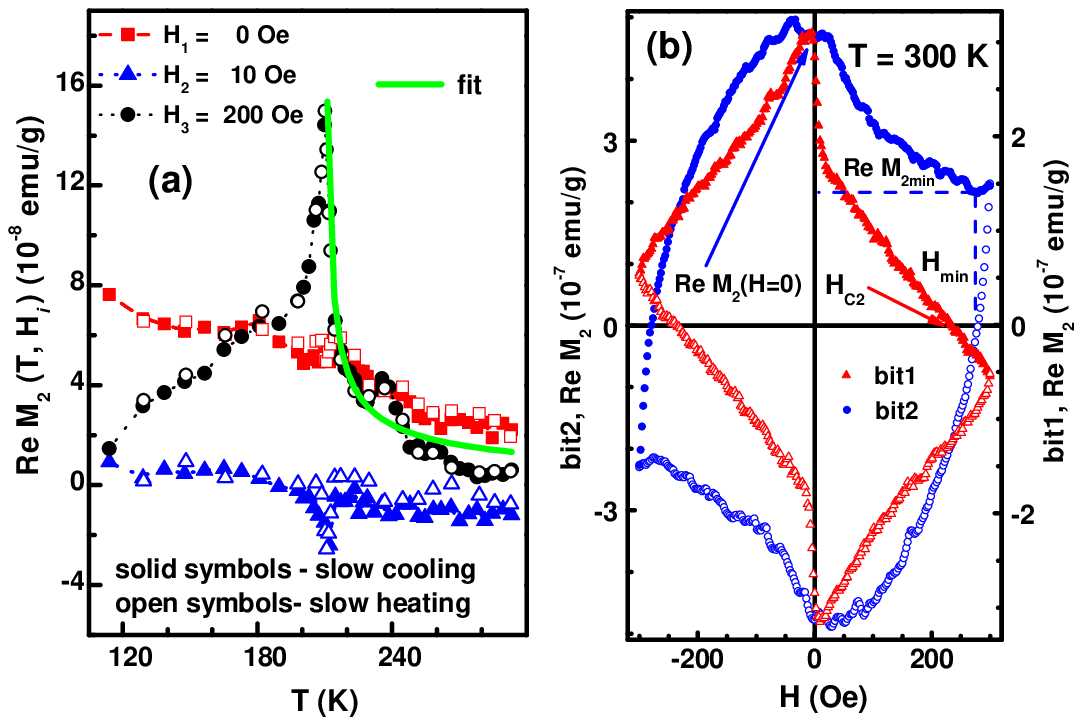}
\textbf{Figure 7.}
\end{center}

\begin{center}
\includegraphics*[width=1.3\linewidth, keepaspectratio=true]{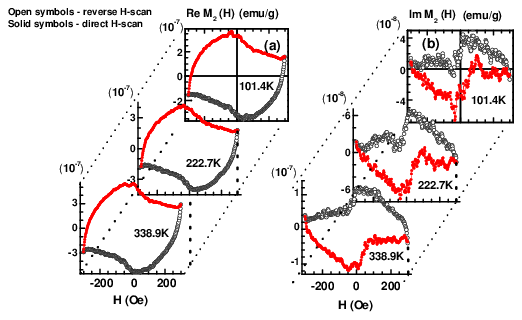}
\textbf{Figure 8.}
\end{center}

\begin{center}
\includegraphics*[width=1.3\linewidth, keepaspectratio=false]{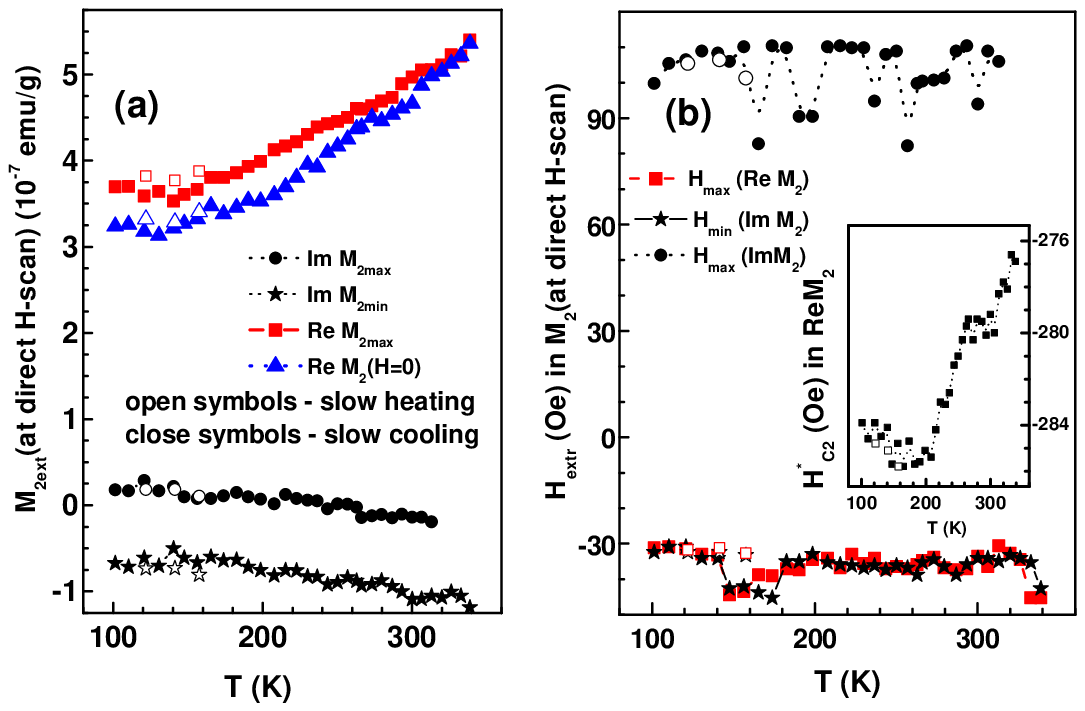}
\textbf{Figure 9.}
\end{center}

\begin{center}
\includegraphics*[width=1.3\linewidth, keepaspectratio=false]{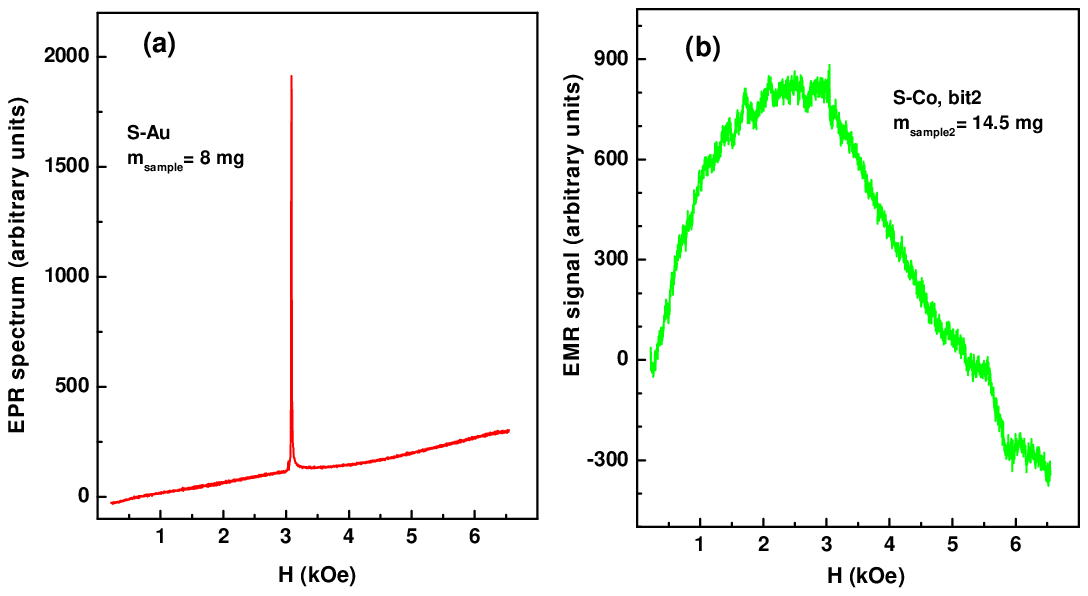}
\textbf{Figure 10.}
\end{center}

\begin{center}
\includegraphics*[width=1.3\linewidth, keepaspectratio=false]{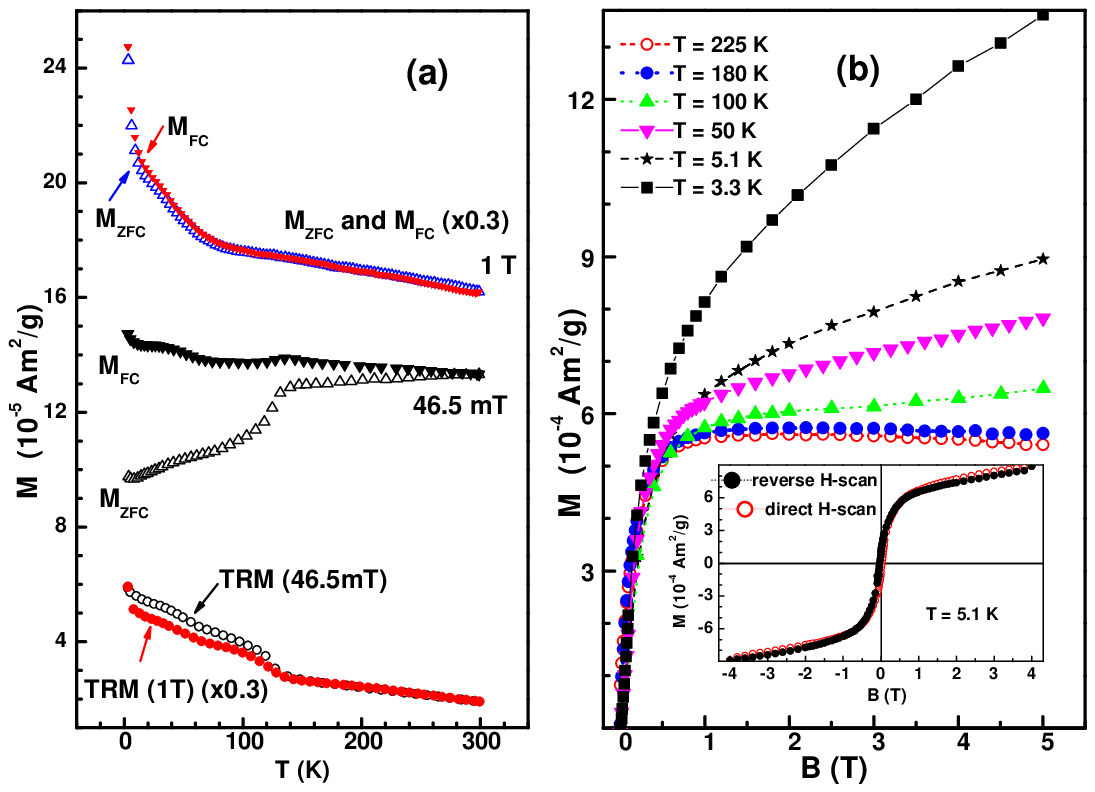}
\textbf{Figure 11.}
\end{center}

\begin{center}
\includegraphics*[width=1.3\linewidth, keepaspectratio=false]{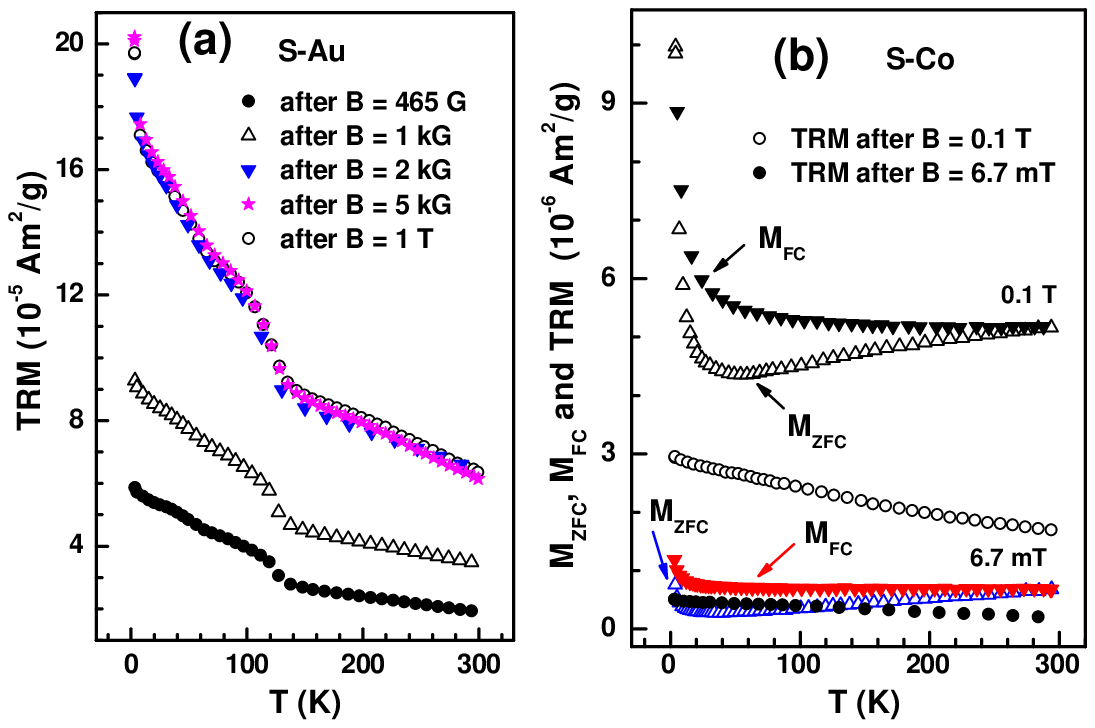}
\textbf{Figure 12.}
\end{center}

\end{document}